\documentclass[runningheads]{llncs}
\input{mystyle.sty}

\begin{document}

	\title{Identifying Emerging Technologies and Leading Companies using Network Dynamics of Patent Clusters: a Cybersecurity Case Study}
	
	\titlerunning{Network Dynamics of Patent Clusters}

    \author{Michael Tsesmelis\inst{1} \and
    Ljiljana Dolamic\inst{1}\and
    Marcus Matthias Keupp\inst{2}\and
    Dimitri Percia David\inst{1,3,4}\and
    Alain Mermoud\inst{1}
    }
    \authorrunning{Michael Tsesmelis et al.}
    %
    \institute{Cyber-Defence Campus, armasuisse Science and Technology\and
    Department of Defense Economics, Military Academy at ETH Zurich\and
    Information Science Institute, University of Geneva\and
    Institute of Entrepreneurship and Management, University of Applied Science (HES-SO) Valais-Wallis
    }
    \maketitle   
	    
	\begin{abstract}
        Strategic decisions rely heavily on non-scientific instrumentation to forecast  emerging technologies and leading companies. Instead, we build a fast quantitative system with a small computational footprint to discover the most important technologies and companies in a given field, using generalisable methods applicable to any industry. With the help of patent data from the US Patent and Trademark Office, we first assign a value to each patent thanks to automated machine learning tools. We then apply network science to track the interaction and evolution of companies and clusters of patents (i.e. technologies) to create rankings for both sets that highlight important or emerging network nodes thanks to five network centrality indices. Finally, we illustrate our system with a case study based on the cybersecurity industry. Our results produce useful insights, for instance by highlighting (i) emerging technologies with a growing mean patent value and cluster size, (ii) the most influential companies in the field and (iii) attractive startups with few but impactful patents. Complementary analysis also provides evidence of decreasing marginal returns of research and development in larger companies in the cybersecurity industry.

	\end{abstract}

\newpage
\section{Introduction}

With accelerating innovation cycles, stakeholders in both industry and government have an increasing need for dependable and real-time insights on technological paradigm shifts. With clear and timely knowledge on emerging technologies within their industries, these stakeholders can adapt the course of their business to accommodate these changes as efficiently as possible and thus reduce costly and unnecessary ventures.

Although data-backed solutions to modern technology monitoring and identification are improving, strategic business decisions still rely heavily on the results of non-scientific instrumentation. Oftentimes, the identification of emerging technologies relies on human intuition rather than robust data science approaches. For instance, consultancy firms continue to use the Delphi method in an effort to predict future business and technology trends, a product that many institutions acquire for the lack of alternatives. This dearth of options cripples many industries, including the cybersecurity sector, which is doubly affected by the poor technology landscape mapping. First, cybersecurity experts should stay up-to-date with innovations within their own field for the industry to perform at its best. Secondly, because the cybersecurity sector monitors and defends other industrial and commercial systems from attack, it is crucial for cybersecurity specialists to understand how other systems and their attack surfaces evolve over time. Thus, the cybersecurity industry is extremely reliant on timely information about emerging technologies and leading companies linked to them.

Hereafter, we present a general recommender system\footnote{ \href{url}{https://github.com/WanderingMike/Network\_dynamics\_of\_patent\_clusters}} suitable for technology monitoring in any industry, which ranks technologies and companies according to useful strategic indicators. The existing literature has mainly focused on ex-post bibliometric analysis, and has done so in very restricted industries. We take a different approach by analysing near-past dynamics and predicting current and near-future trends. This renders our solution much more useful for applied technology monitoring. 
For such a data-driven system to run dependably over long stretches of time, in organisations with limited computational and human resources, we recognise the need for four major features. First, as much of the new solution should be automated to allow efficient and error-free data search and processing. The second feature, closely linked to the previous one, is speed. Since open source data is usually published several times yearly, it is preferable to run more lightweight algorithms at shorter intervals rather than a long-running algorithm. Thirdly, the solution must not be overreliant on any one indicator, but should rather take the financial approach of comparing a multitude of indicators before making a selection of which companies or technologies to highlight. This feature also inscribes itself well in the theory of artificial intelligence, where decision-making is the product of many different data sources and knowledge representations. Finally, the solution must be generalisable, thus allowing different search criteria (based on different technological fields). This factor presupposes a system reactive to paradigm changes across time and industries.

This paper will describe the design of a system that identifies rising and impactful technologies and the companies tied to them. The solution will rely on a sequential blend of machine learning and network analytics and will be fed data from the United States Patent Office (USPTO). Our research highlights how a lightweight data-driven system can produce useful insights. Our cybersecurity case study shows clear and fine-grained technology rankings which point to rising interest in neural networks, payment technologies,  control units, WiFi security and cryptography. The system also successfully identifies the most impactful big technology firms as well as pre-eminent startups in the field. This information provides useful information for systems acquisition, equity investments, or technology policy.

\section{Related Work}

While the meaning of the word 'emerging' seems intuitive, it is difficult to recognise a commonly accepted definition which can be used to label technologies in our recommender system. A large number of publications attempt to define the concept, which has resulted in a convolution of definitions which sometimes overlap and in certain regards differ widely. For some authors \cite{porter2002, Martin1995, Corrocher2003, Hung2006, Halaweh2013}, a technology is deemed emerging when its potential impact on the economy or society is high, a terminology which includes both evolutionary changes as well as disruptive innovations. Other scholars \cite{Boon2008} pay more attention to the intrinsic uncertainty of a technology's rise. Some deem it a combination of both of these aspects that affect how "emerging" a technology is. Another group \cite{Small2014} underlines novelty and growth as key determinant factors. Finally, the technological convergence literature believes that emerging technologies arise when two originally independent technologies fuse to give rise to a technological breakthrough \cite{wang_2019}.

This failure to approach a reasonably clear definition has naturally played to the advantage of qualitative methods of investigation. Some institutions, such as Gartner, Forrester Research, IHS Markit and the World Economic Forum regularly publish lists of emerging technologies with very little indication on how these lists are curated. Despite differences both in terms of frequency of publication and length, the contents of these lists often agree, thus pointing to a certain statistical significance of the published recommendations.

On the other hand, quantitative studies of emerging technologies generally agree that measuring a proxy of growth and technology interest provides enough information for industrial applications. Indeed, even though it might be hard to define whether one particular technology is emerging, it is much easier to compute the relative importance of technologies. The entire analysis thus becomes a comparison between different technologies using common metrics in order to differentiate rising from relatively stale or declining technologies. We follow this train of thought in building our own system, and regard emerging technologies as those with the highest growth rates in citations and patent count. In our framework, emerging technologies are therefore not necessarily new, but witness the fastest increases in interest from researchers and patent applicants. 

Publicly available data is often used to predict emerging technologies in the existing literature. Commonly exploited data sets are patents from the United States Patent and Trademark Office (USPTO), Global Patent Index (GPI), and Thompson Innovation. Some publications use these bibliometric sources to deploy growth models for technology prediction \cite{andersen1999, Intepe2012, Nieto1998, Ranaei2014}. S-Curve models are based on the concept of logistic or Gompertz growth that eventually leads to saturation. The clear advantage of such models is their sound mathematical foundation. After fitting the data to the model, the exact growth formula displays the current maturity state of a technology and its future development potential. One research study \cite{Daim2006} applied bibliometric methods, US patent analysis and S-curves for forecasting fuel cells, food safety, and optical storage using a technology-specific set of methods. Similarly, \cite{Ranaei2014} used expert interviews to fit data acquired by text-mining patents into growth curve models for predicting hybrid cars and fuel cells. Text-mining on patents and fitting to S-curves was also proposed in \cite{Kucharavy2009}, and \cite{Bengisu2006} found a correlation between patent and publication data extracted by scientometric methods for 20 technologies and deployed S-curves for forecasting. S-Curve models for predicting emerging technologies were also proposed by \cite{Intepe2012, Nieto1998}.

Recently, artificial intelligence has (re-)gained much attention and consequently machine learning has been deployed to model and predict emerging technologies. \cite{Kyebambe2017} used supervised learning on citation graphs from USPTO data to automatically label and forecast emerging technologies with high precision in a given year. Similarly, \cite{Zhou2018} applied supervised deep learning on worldwide patent data. The training sets were labelled manually based on Gartner's Hype Cycle. \cite{Lee2018} extracted 21 indicators from the USPTO data and using neural networks achieved impressive predictive power on a subset of 35,256 patents. The big issue with these studies is their focus on a specific set of technologies chosen in advance \cite{Intepe2012,Nieto1998, Ranaei2014}. As a consequence, most recommender systems can only be applied to a new case study after a painstaking process of data curation and machine learning model selection. This calibration can require the full-time attention of a small team of data scientists.

We thus recognise that the existing literature shows good results in applying novel statistical and computational methods to analysing and predicting the success of technologies within a specific industry. However, our goal is rather to pursue a lightweight system that can easily switch between different user queries, with fast results. This would in effect solve the lack of quantitative options in technology monitoring that we highlighted previously by offering a reliable alternative applicable to any industry with no additional effort. Moreover, we recognise the need to involve not just technologies into our system, but also interesting companies tied to these technologies. One strand of research by \cite{mezzetti2021} has studied the mutual influence of companies and technologies from the cybersecurity field using bi-partite graph structure. We use such an approach to create networks of technologies and companies in our own recommender system.

We do rely on existing solutions to optimise the speed and efficiency of our solution. Our system deals with resource constraints and incomplete data sets by using probabilistic models in machine learning to simplify the research problem and thus its computational complexity; rather than using the entire data set of all artifacts (whether it be patents, publications, job openings, and so on), we reduce the space complexity of the problem by working with distributions of data points described by simpler statistics such as the mean and the variance. Financial engineering also offers an improved approach to high-variance problems; rather than using a single forecasting method, the field suggests averaging the forecasts obtained thanks to several instruments \cite{makridakis_1983, clemen_1989}. This approach attenuates the variance on the final results and thus offers more dependable insights. In our system, this translates to using several indicators and data sources to confirm the importance of particular technologies and companies. With these methods in place, we believe our system offers a novel way of ranking emerging technologies and leading companies. 
\section{Data sources}
\begin{figure}[h]
    \includegraphics[width=1\textwidth]{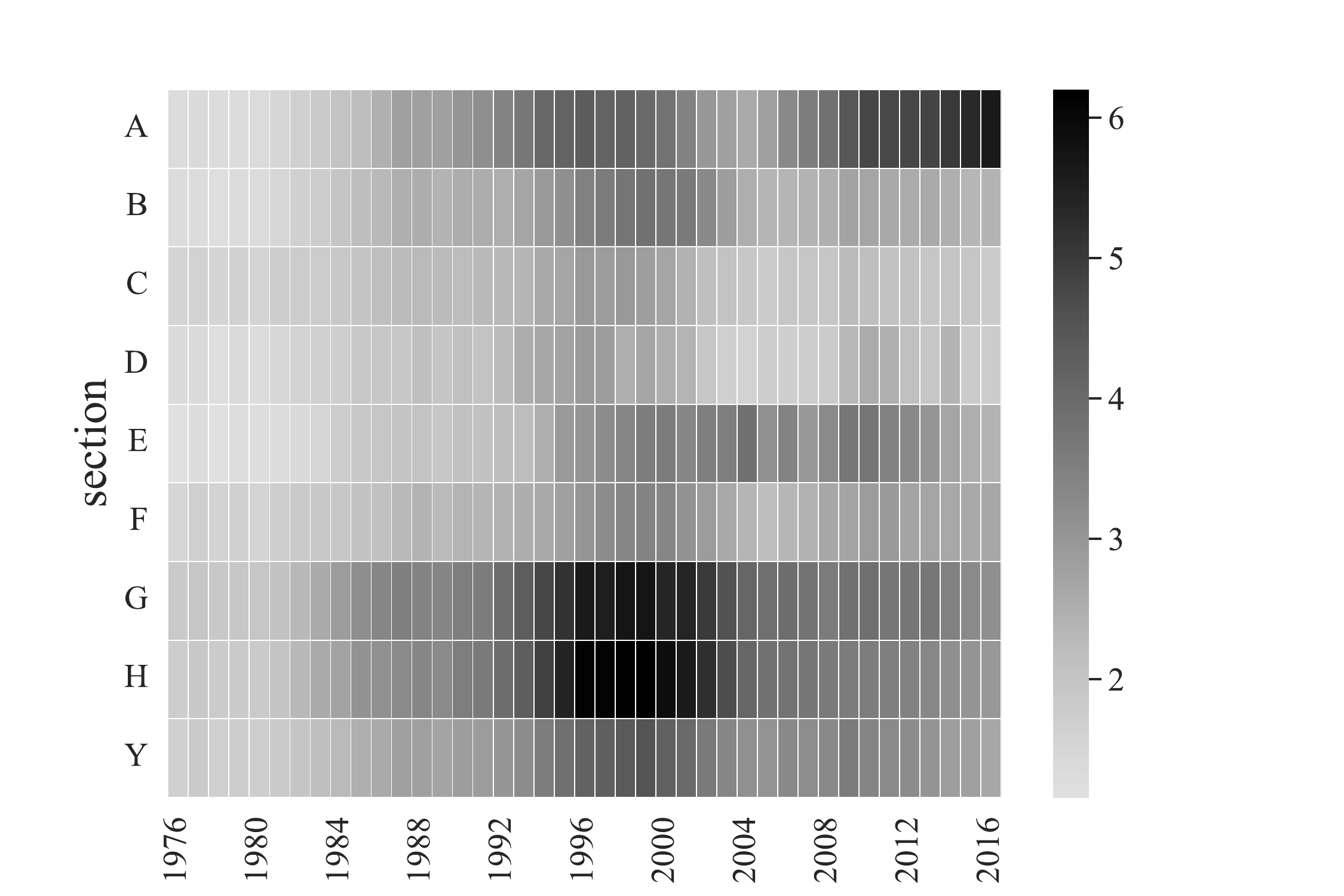}
    \caption{Average five-year forward citation value of the nine different Cooperative Patent Classification technology sections. All patents belong to at least one section. \newline A - Human necessities (agriculture, garments,...); B - Performing Operations \& Transporting; C - Chemistry \& Metallurgy; D - Textiles \& Paper; E - Fixed Constructions; F - Mechanical Engineering, Lightning, Heating, Weapons \& Blasting; G - Physics; H - Electricity; Y - New technological developments}
    \label{fig1}
\end{figure}

We build our recommender system using patent data from the US Patent and Trademark Office (USPTO). Patents provide an essential means to capture the growth trajectory and novelty of a technology \cite{andersen1999, meyer2001patent, haupt2007patent}. In order to determine the importance of a new technology cluster, it is however not enough to simply count the number of patents related to a technology, as this ignores the relative value of each patent. Instead, we decide to assign a value to each patent based on the number of citations it receives in the five-year period after its publication. This indicator, termed the five-year forward citation (5YFC) value, is the basis for the Machine Learning algorithm. We have chosen a five-year horizon as a benchmark, since the literature estimates the median forward citation to occur between the fourth and fifth year after a patent's publication \cite{Lee2018}. Figure \ref{fig1} shows the average 5FYC value for different fundamental science categories per year. For instance, information technology and physics innovations of the late 1990s were highly cited in the five years following their publications.

Broadly speaking, our research will analyse the interplay of three sets of entities. Firstly, data on patents $p$ help us differentiate valuable from non-valuable patents. Secondly, each patent is usually sponsored by one or more assignees $a$, which are organisations or individuals that have an ownership interest in the patent claims. Assignees are added to the network as separate entities, and thus we retrieve assignee and assignee-patent relationship data from the USPTO database. Finally, based on their protected claims, patents are tied to specific Cooperative Patent Classification (CPC) groups $c$. CPC is a classification scheme which clusters patents according to the subject matter of the patented innovation. CPC, along with the US and International classification schemes, allow innovations to be grouped into predefined technological clusters. The overall hierarchy scheme of the CPC is as follows: $Subgroup \subseteq Group \subseteq Subclass \subseteq Class \subseteq Section$. We rely on the 242,050 CPC subgroups to define all currently existing technologies, and retain data on groups for the machine learning phase. All downloaded data sets are found in Table \ref{Table1}. 

\begin{table*}[h]
    \footnotesize
    \centering
    \begin{tabular}{p{0.3\textwidth}p{0.7\textwidth}}
    \toprule
    \textbf{Data set name} &  \textbf{Description} \\
    \midrule
    \textbf{assignee.tsv}           & Disambiguated assignee data for granted patents and pre-granted applications, of which we retain the assignee ID and the company name. It is important to note the geographical diversity of the assignees; many non-American companies have submitted documentation to protect their inventions in the United States, and thus our data source extends beyond the confines of the American technological landscape.\\
    \textbf{cpc\_current.tsv }      & Current CPC classification data for all patents, of which we retain the patent ID, CPC group ID and CPC subgroup ID.\\
    \textbf{otherreference.tsv}     & Non-patent citations mentioned in patents (e.g. articles, papers, etc. We only retain the patent ID column in order to count the number of references each patent makes to non-patent literature.\\
    \textbf{patent.tsv}             & Data on granted patents, of which we retain the patent ID, the grant date, the abstract text, and the number of claims.\\
    \textbf{patent\_assignee.tsv}   & Metadata table for many-to-many relationships between assignees and patents, of which we retain the patent ID and assignee ID.\\
    \textbf{patent\_inventor.tsv}   & Metadata table for many-to-many relationships between patents and inventors, of which we retain only the patent ID information in order to count the number of inventors present on application documents.\\
    \textbf{uspatentcitation.tsv}   & Data on citations between US patents, of which we retain the cited patent ID and the citing patent ID. \\
    
    \bottomrule
    \end{tabular}
    \vspace{0.5cm}
    \caption{Full list of required data sets. Their content is described as well as the retained data points, which appear as columns in a tab-separated format. The data sets can be downloaded from \href{https://patentsview.org/download/data-download-tables}{Patentsview}.}
    \label{Table1}
\end{table*}

Patentsview provides patent information from the United States Patent Office (USPTO) for all patents issued since 1976. The bulk data sets are available for download on their website. We relied on the data set available in October 2021 which contains 7'101'932 patents. All data sets are open to the public and available free of charge. Additionally, data sets are regularly maintained and thus represent an up-to-date repository of scientific evidence. 

\section{Methodology}
Our recommender system consists of four layers, shown in Figure \ref{fig2} below. First, patent data is cleaned in the Data Preprocessing layer. Secondly, in the Machine Learning layer, we cluster these patents and extract key descriptive features to train machine learning classifiers. The output of the Machine Learning layer is a labeled set of patents, split between low- and high-value patents. The Managerial layer then selects relevant patents based on the user's query and passes them further on. The Network Analytics layer receives these topical patents and builds a graph of patent clusters (i.e. technologies) and patent assignees (i.e. companies), which our system uses to calculate useful indicators and rankings of both sets of graph nodes.

\begin{figure}
    \figuretitle{Layered system flowchart}
    \includegraphics[width=1\textwidth]{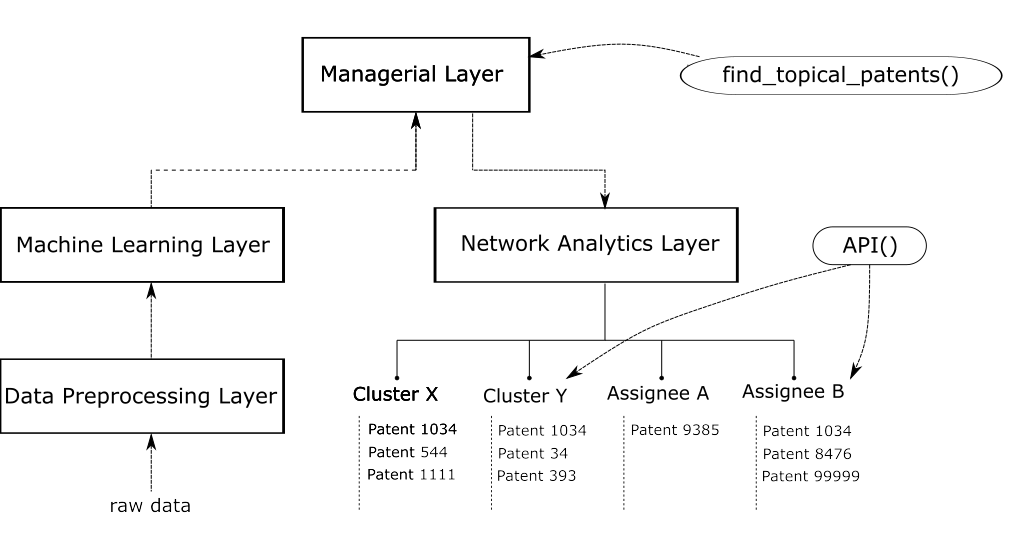}
    \caption{Flowchart depicting the four different layers of the system as well as the data flow through it. The function find\_topical\_patents\(\) finds overlaps between patents abstracts and the keyword input given by the user. The API allows users to investigate the network of clusters and assignees created by the Network Analytics layer.}
    \label{fig2}
\end{figure}

\subsection{Data Preprocessing layer}
The Data Preprocessing layer is the intersection between the USPTO database on the outside and our system on the inside. As such, it plays a vital role in cleaning and preprocessing the data in a way that satisfies the requirements of the algorithm working downstream. A successful Data Preprocessing layer performs three tasks: (i) trimming of data sets to the required information, (ii) preventing easy and repeatable errors early on, and (iii) producing lightweight data containers.

To satisfy (i), each data set is loaded and unnecessary data columns are removed; the remaining ones are relabeled with intuitive and standardised names used across the system scripts. This work is also a forward-thinking process, as it entails determining the data prerequisites of the Machine Learning and Network Analytics layers early on. The Machine Learning layer requires descriptive information about patents to forecast their 5YFC. On the other hand, the Network Analytics layer needs many-to-many tables to assign patents $p$ their respective assignees $a$ and CPC subgroups $c$.

For task (ii), freeing the program from easy errors should be seen as a commitment to understanding the data sets. Each column is assigned a data type (e.g. integer, string, dates, etc) and missing values are updated with common replacements such as the median of the column, a plain 0, or NaN values.

Finally, the approach of our algorithm to task (iii) allows us to compute clusters and networks much faster and with less memory overhead. Indeed, common practice would load all data sets as data frames in random-access memory (RAM), a space-hungry method. With such large arrays, it is not uncommon to witness memory shortages on consumer-grade computers with 16- or 32GB RAM space. Creating a more lightweight algorithm required thinking beyond the traditional data science methods. Options such as creating an external SQL database were too slow, and reading data sets sequentially was prone to errors. We opted for storing the data using multi-dimensional dictionaries (hereafter called tensors), saved locally using pickle file formatting rather than the heavier comma-separated-value format.

\subsection{Machine Learning layer}
In studies such as \cite{Lee2018}, accuracy scores on classification problems using patents remain low, especially for time frames beyond 2 years. Additionally and more dangerously, the highly unbalanced training data sets of such studies skew the results in favour of the dominant class, and hence the algorithm's overall performance is kept artificially high. We opt for a more robust approach by first applying a statistical analysis to the data set and based on these results adapting our output classes. Instead of four possible classification values as found in the study of \cite{Lee2018}, we reduce our problem to a binary classification system with a patent value $value_p$ of 0 for low-value patents and 1 for high-value patents. In order to label each patent, we first measure the distribution of the 5YFC values of patents older than 5 years. We then define high value patents as those with a 5YFC above the third quartile. Following this, we train the machine learning algorithms on an identical number of low- and high-value patents. Moreover, in order to improve the applicability and robustness of our algorithms across patents from all 242,050 CPC subgroups, we take for our classification system a randomised selection of patents. To boost the Machine Learning layer further, we simplify the input indicators fed to the algorithm and test our training set on a wide selection of classification algorithms.

 The 5FYC can easily be gleaned for patents older than 5 years, since our data sets contain that information. However, these values are missing for the most recently published patents. It is precisely to fill this gap that we deploy machine learning methods. We first compile a shortlist of input variables, most of them copied directly from \cite{Lee2018}. Table \ref{Table3} summarises these indicators. A total of 13 indicators are used as input data for a supervised classification problem with the output value $value_p$. Rather than using a single machine learning algorithm, we search for the best-in-class by testing a wide sample of general classifiers on our data set and comparing performance levels. In this task, we are supported by the auto-sklearn framework, which automatically selects the best possible model and calibrates its hyperparameters to maximise classification scores with patents that already have a 5YFC value. 

\begin{table*}[h]
    \footnotesize
    \centering
    \begin{tabular}{p{0.50\textwidth}p{0.50\textwidth}}
    \toprule
    \textbf{Indicator} &  \textbf{Description} \\
    \midrule
    Five-year forward citation (5YFC)           & Binary value representing the value of a patent, 1 being high-value, 0 low-value\\
    Class-level technological originality (CTO) & Herfindahl index on CPC groups of cited patents    \\
    Prior knowledge (PK)                        & Number of backward citations    \\
    Scientific knowledge (SK)                   & Number of non-patent literature references \\
    Technology cycle time (TCT)                 & Median age of cited patents \\
    Main field (MF)                             & Main class to which a patent belongs \\
    Technological scope (TS)                    & Number of classes to which a patent belongs \\
    Protection coverage (PCD)                   & Number of claims \\
    Collaboration (COL)                         & 1 if a patent has more than one assignee, else 0 \\
    Inventors (INV)                             & Number of inventors \\
    Total know-how (TKH)                        & Total number of patents issued by an assignee \\
    Core area know-how (CKH)                    & Number of patents in a CPC group of interest issued by an assignee \\
    Total technological strength (TTS)          & Number of forward citations of patents issued by an assignee \\
    Core technological strength (CTS)           & Number of forward citations of patents in a CPC group of interest issued by an  assignee \\
    \bottomrule
    \end{tabular}
    \vspace{0.5cm}
    \caption{Summary of patent indicators used by the machine learning algorithms. The 5YFC is the binary output variable, the rest are input variables.}
    \label{Table3}
\end{table*}

Once the training phase is over, we predict the binary value of the most recently published patents. Finally, for each cluster $c$ in year $n$, we measure the mean value of its patents as well as the yearly patent count. Thus, we get:
\newpage

\begin{equation}
size_{c, n} = |c_{n}|
\end{equation}

\begin{equation}
value_{c, n} = \frac{1}{size_{c, n}}\sum_{value_{p} \in c_{n}}value_{p}
\end{equation}

\noindent
where
\begin{itemize}
 \item[] $c_n$ is a technology cluster belonging to $C$ and containing patents issued in year $n$
 \item[] $value_p$ is the patent value (0 or 1) for patent $p$
\end{itemize}

\subsection{Managerial layer}
The Managerial layer is the front end of our system and deals with user input in the form of jobs. These jobs contain settings to adjust job duration and to specify job content. By giving the system a list of unique or concatenated keywords related to a search topic, the user designates a particular area of interest. The managerial layer receives these keywords and scans all 7'101'932 patent abstracts for occurrences of these words. The patents in which these concepts appear verbatim are saved in a list of topical patents. Each job thus produces a list of patents strongly or weakly connected to the technological query. From the managerial layer, we then activate for each job and list of topical patents a network that will map technologies and companies related to these patents. 

Although our approach to identifying patents related to specific queries can be deemed elementary, we believe this approach results in a more comprehensive and statistically significant view of the patent corpus. By specifying a keyword list that is both precise - i.e. bearing little overlap with unrelated queries - and complete - i.e. capturing all essential traits of the search -, the abstract search module captures most existing connections between the search query and the patent corpus without the need for complex mechanisms that often result in false positives. In the Network Analytics layer, each patent will lead us to its CPC subgroups, and this voting mechanism will activate some CPC subgroups more than others. Thus, CPC subgroups that deal heavily with the terms of our queries will outweigh nodes that are only sparingly mentioned. This correction mechanism underlines the robustness of investigating large data sets instead of focusing on fewer data points, which is often the case in the previous literature on technology forecasting.

\subsection{Network Analytics layer}
The Network Analytics layer is the final processing layer and is responsible for computing different indicators that measure technology emergingness and company rankings. On the basis of the queries made in the managerial layer, the Network Analytics layer selects relevant CPC subgroups (also known as topical clusters) associated with the topical patents. We permeate this step further by first retrieving all patents belonging to these selected topical clusters, and secondly all assignees that sponsored these latter patents (hereafter named topical assignees). From this sequential process, we thus end up with two sets of entities from which we build a graph. A graph is a structure used to model pairwise relationships between entities \cite{mezzetti2021}. In our solution, these entities are on the one hand a subset of the 242'050 CPC subgroups $C$, with each topical cluster $c$ forming a separate node in the graph. The second set of topical assignees $A$ is directly associated with these CPC subgroups, and each assignee $a$ forms a separate graph node as well. We specifically build a bipartite network, whose nodes can be divided into two disjoint sets $C$ and $A$, with all edges $E$ between nodes joining one node of each set and never two entities of the same set. 

Since our research seeks to highlight the latest emerging trends in technologies, we use only the most recent data in the time-series produced by the Machine Learning layer to measure emergingness values. Across the set $N$ of all years covered by the data set, we measure technology emergingness as the summation of three growth factors between the years $argmax\;N - 3$ and $argmax\;N$. This step considerably reduces the complexity of the computational problem: rather than working with a long time-series of graphs, we only require four year's worth of patents to identify emerging nodes. Subsequently, each CPC subgroup node is weighted by its value $value_{c,n}$ and labeled with its patent count $count_{c,n}$. Secondly, the edges $e \in E$ between entities are weighted by the amount of patents that a tuple ($c\in C$, $a \in A$) have in common.

We decided to build 5 indices to measure node centrality, which highlight important technologies and companies based on different criteria. The values of these indices all relate to the specific job query.

\subsubsection{Technology Index}
The Technology Index is the only index of our methodology pertaining to technologies rather than to companies. In chapter 2, we described several concepts used in defining emerging technologies. We hereafter choose to define emergingness as an increase in overall interest and impact of a technology across time. In our system, we have defined each of 242,050 CPC subgroups as potential technologies. The absolute interest in these technologies is hereby proxied by the number of patents granted by the USPTO in each CPC subgroup per year, i.e. $count_{c,n}$. The impact measure on the other hand is proxied by $value_{c,n}$. Then, to calculate the emergingness of a technology, we consider the difference of these two values between year $n$ and $n-1$ in the range between $argmax\;N - 3$ to $argmax\;N$. Thus, we have:

\begin{equation}
value\:growth_{c, n} = \frac{value_{c,n}}{value_{c,n-1}}
\end{equation}

\begin{equation}
cluster\:growth_{c, n} = \frac{size_{c,n}}{size_{c,n-1}} \times \sqrt[m]{size_{c,n-1}}
\end{equation}

\begin{equation}
tech\:index_c = \frac{1}{3}\sum_{n \in N}{value\:growth_{c,n} \times cluster\:growth_{c,n}}
\end{equation}
where
\begin{itemize}
 \item[] $ m $  calibrates penalty for small patent clusters (here, $m$=5)
 \item[] $ c $  one technology cluster (CPC subgroup cluster) in $C$
 \item[] $ n $  one year between $argmax\;N - 3$ to $argmax\;N$  
\end{itemize}

\subsubsection{Assignee Quality Index}
Given our query, we have built a topology of relevant assignees, called topical assignees. These have in common at least one granted patent in a set of query-relevant technologies. This index is obtained by measuring the mean value of the patents assigned to assignee $a$ in year $argmax\;N$. This index highlights  research institutions producing the highest quality or impactful research. We thus have:

\begin{equation}
value_{a} = \frac{1}{size_a}\sum_{value_{p} \in a}value_{p}
\end{equation}

\subsubsection{Impact Index}
Rather than looking at the value of the patents an assignee has submitted, the Impact Index measures the value of the contribution of an assignee to a query and rewards assignees that have strong connections to important technologies in the network. We define the Impact Index as a sum of the relationships each assignee has with all topical clusters in the network. Given an assignee $a$ and a CPC subgroup $c$, the strength of this relationship is defined as the product of the $value_a$ of the assignee node, the $value_{c, n}$ of the CPC subgroup node, and the network edge weight between the two entities $weight_{e}$. This index captures the overall contribution of a company to a field, and hence highlights companies foundational to the area of interest. We have:

\begin{equation}
weight_{e} = |\{\{c,a\}\;|\;c \in C,\:a \in A\}|
\end{equation}

\begin{equation}
impact_{a} = \sum_{e \in E_a} value_a \times value_{c, \argmax N} \times weight_{e}
\end{equation}

\noindent
where
\begin{itemize}
 \item[] $E_a$ is the set of all edges \textit{e} connected to assignee \textit{a}
\end{itemize}

\subsubsection{Normalised Impact Index}
The Normalised Impact Index is a proportionally weighted version of the Impact Index. Indeed, we recognise that some smaller and lesser known companies with a high proportion of high-value patents will not appear prominently in the Impact Index due to their small number of patents. The Normalised Impact Index corrects for this oversight. This index is particularly prone to highlighting influential startups. We have:

\begin{equation}
norm index_{a} = \frac{impact_a}{\sum_{e \in E_a} weight_e}
\end{equation}

\subsubsection{Influence Index}
In graph analytics, a common metric of network centrality is obtained with the eigenvector centrality. Discussed extensively in \cite{Ruhnau_2000}, eigenvector centrality computes the centrality for a node based on the centrality of its neighbours. Thus, this index highlights well connected and influential companies in the network. The eigenvector centrality for node \textit{i} is the \textit{i}-th element of the vector \textit{x} defined by the equation:

\begin{equation}
Ax = \lambda x
\end{equation}

\noindent
where
\begin{itemize}
 \item[] $A$ is the adjacency matrix of the Graph G with eigenvalue $\lambda$
\end{itemize}

\noindent
We get two measures, the influence of assignee nodes $influence_a$ and that of CPC subgroup nodes $influence_c$, of which we retain only the first.

\section{Empirical analysis and results}

Our initial motivation for building a robust and lightweight technology monitoring system was to encourage anticipation efforts in the cybersecurity sector. To illustrate the performance of our system in those settings, we ran a job using a manually curated list of keywords capturing the essence of cybersecurity. These keywords were chosen from the cybersecurity glossary published by the National Initiative for Cybersecurity Careers and Studies (NICCS) and were retained for their specificity, i.e. each word pertained only to cybersecurity-related topics. We have selected the following words: \newline
'allowlist',
'antimalware',
'antispyware',
'antivirus',
'asymmetric key',
'attack signature',
'blocklist',
'blue team',
'bot',
'botnet',
'bug',
'ciphertext',
'computer forensics',
'computer security incident',
'computer virus',
'computer worm',
'cryptanalysis',
'cryptography',
'cryptographic',
'cryptology',
'cyber incident',
'cybersecurity',
'cyber security',
'cyberspace',
'cyber threat intelligence',
'data breach',
'data leakage',
'data theft',
'decrypt',
'decrypted',
'decryption',
'denial of service',
'digital forensics',
'digital signature',
'encrypt',
'encrypted',
'encryption',
'firewall',
'hacker',
'hashing',
'keylogger',
'malware',
'malicious code',
'network resilience',
'password',
'pen test',
'pentest',
'phishing',
'private key',
'public key',
'red team',
'rootkit',
'spoofing',
'spyware',
'symmetric key',
'systems security analysis',
'threat actor',
'trojan',
'white team'.

We ran our job on a specialised high-performance computer (HPC) node with 128 AMD EPYC 7742 CPUs. We however estimate 16 CPUs to suffice for such a task. The maximum concurrent RAM usage stands at around 100GB, half of which is occupied by the ten different tensors loaded with Patentsview data and the rest occupied intermittently by the large machine learning data frames. Execution time is highly correlated with the machine learning step size. We ran a best-model search using the auto-sklearn framework, capped at 0.6 hours testing time per model and 6 hours in total. The training set consisted of 15,000 low-value patents (binary value 0) and 15,000 high-value patents (binary value 1), thus consisting of a perfectly balanced data frame with a random guess accuracy rate of 50\%. The auto-sklearn framework ran 37 different target algorithms. One algorithm crashed, 3 exceeded the set time limit, and 5 exceeded the memory limit. In the end, the algorithm opted for a mix of differently parametrised random forest models, attaining an accuracy value of 0.6737, or about two in three patents. Following this, we predicted the output values of the 1'740'613 patents granted in the latest five-year period. This step required over 11 hours. All in all, the Data Preprocessing and Machine Learning layers required approximately 26 hours to run. In the Network Analytics layer, different jobs can run successively on top of these results; with $n$ the amount of topical patents, each job has an approximate time complexity of $O(\log{}n)$, since topical patents usually share topical clusters. 

\begin{figure}
    \figuretitle{Cumulative distribution functions (CDF) of clusters}
    \centering     
    \subfigure[Figure A]{\label{fig3:a}\includegraphics[width=60mm]{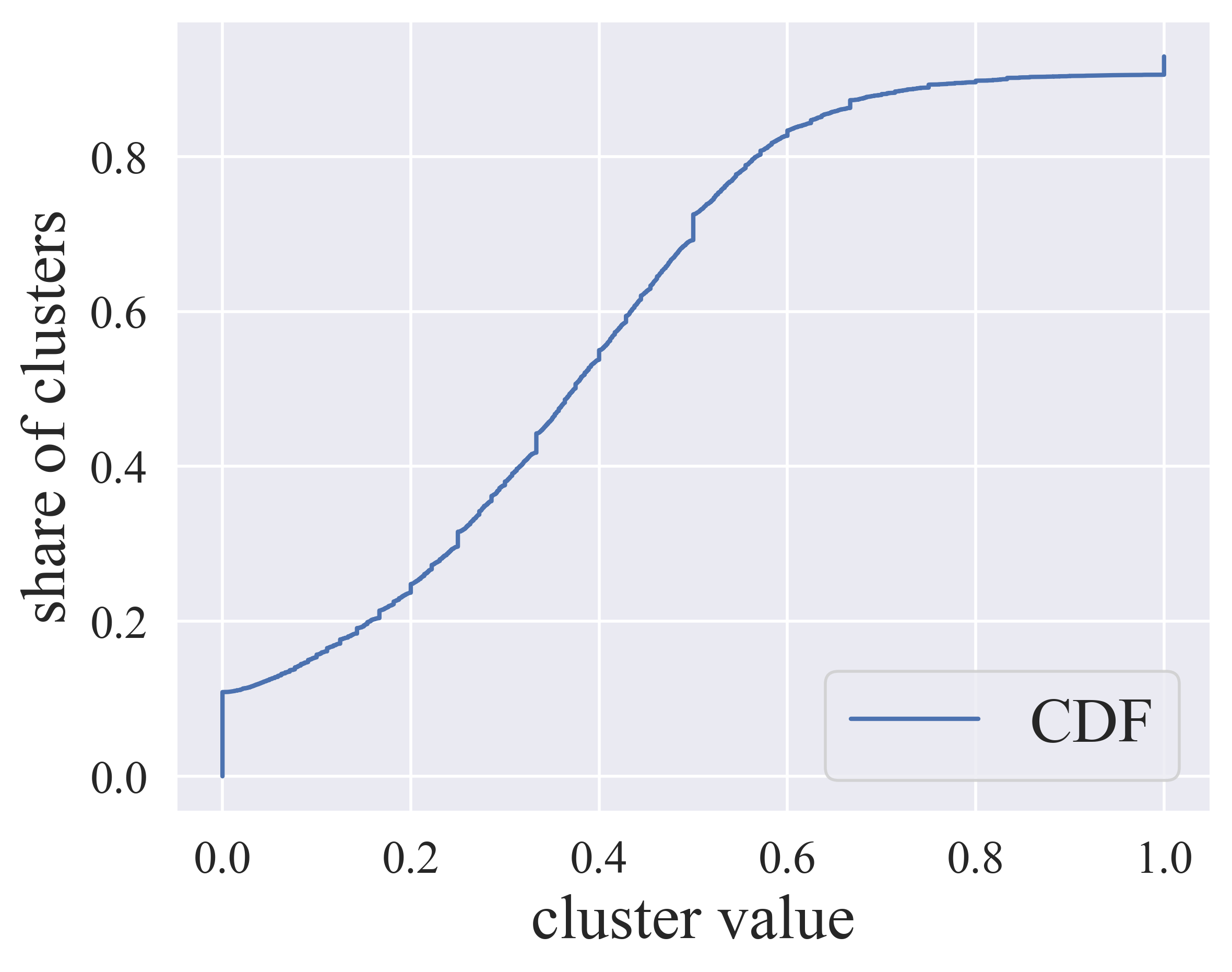}}
    \subfigure[Figure B.]{\label{fig3:b}\includegraphics[width=60mm]{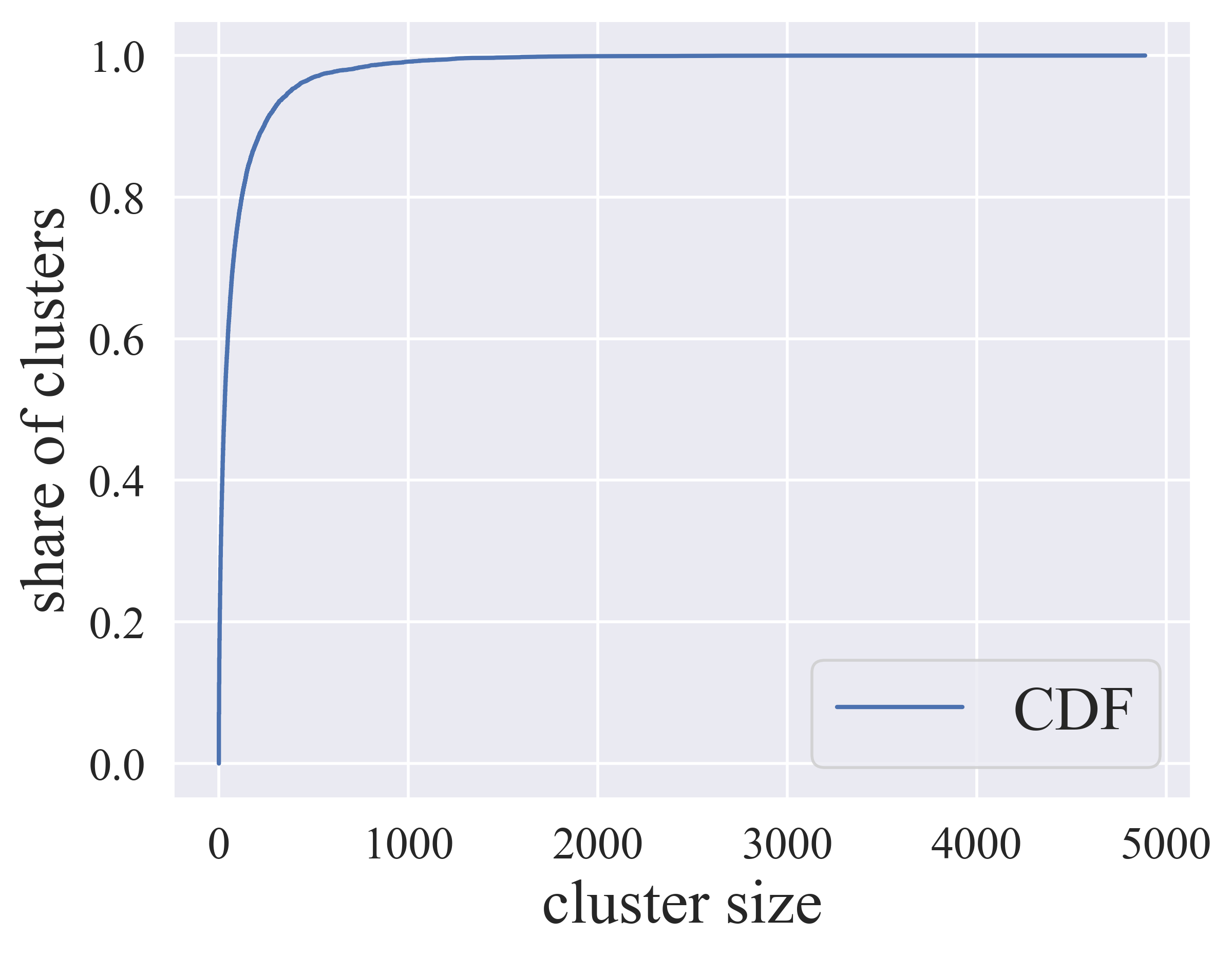}}
    \caption{Cumulative distribution functions (CDF) of the cluster value and cluster size for topical clusters. We see from Figure A that most clusters related to cybersecurity have a value between 0.2 and 0.6, meaning they contain a relatively high proportion of high-value to low-value patents. Figure B points to a highly granular cybersecurity environment, consisting of many small technology clusters.}
\end{figure}

\begin{figure}
    \figuretitle{Violin plots for cluster size}
    \includegraphics[width=1\textwidth]{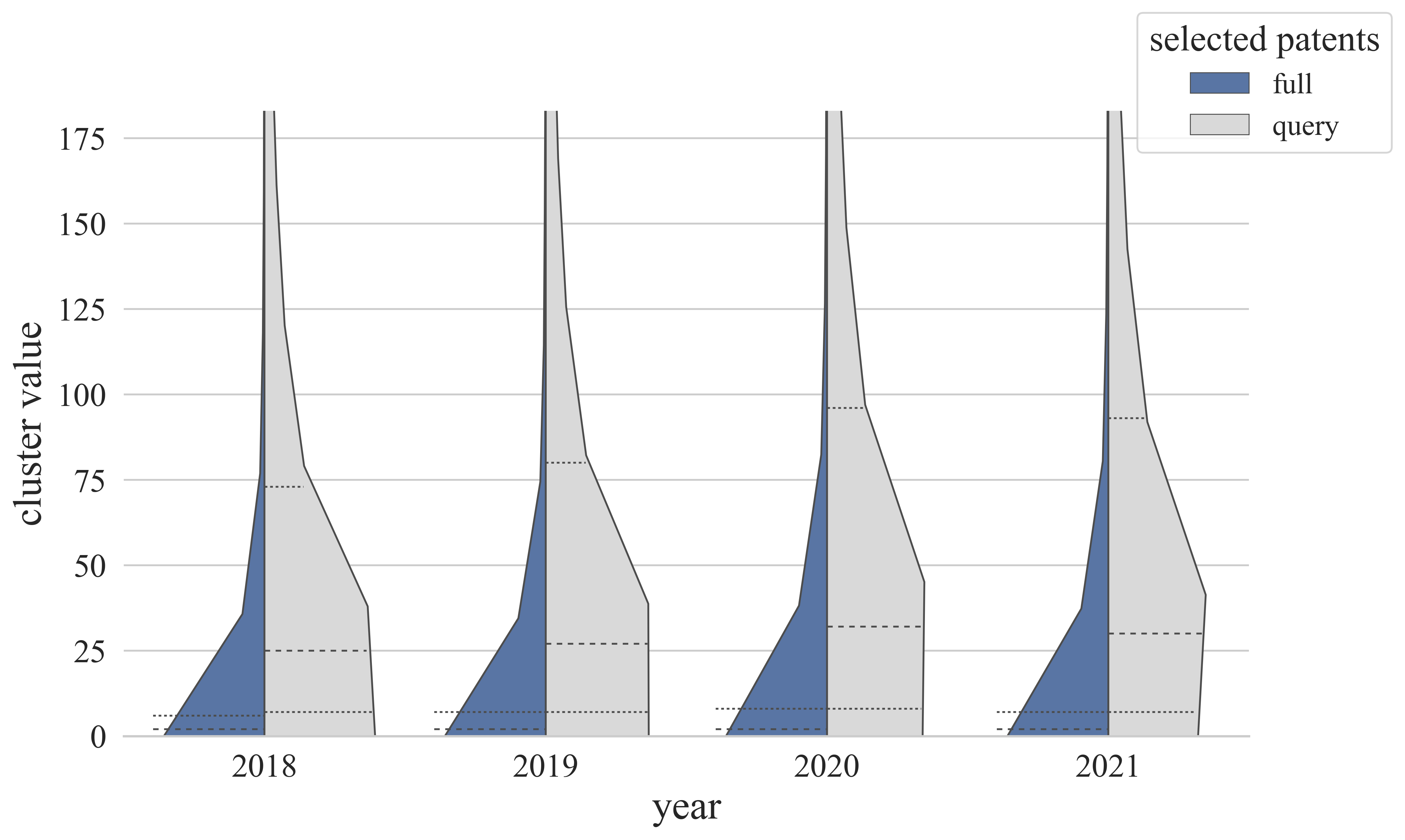}
    \caption{Violin plots for cluster size, from 2018 to 2021. The dashed lines represent the separations between the four quartiles. We see that clusters linked to cybersecurity are bigger than the average cluster.}
    \label{fig4}
\end{figure}

\begin{figure}
    \figuretitle{Violin plots for cluster value}
    \includegraphics[width=1\textwidth]{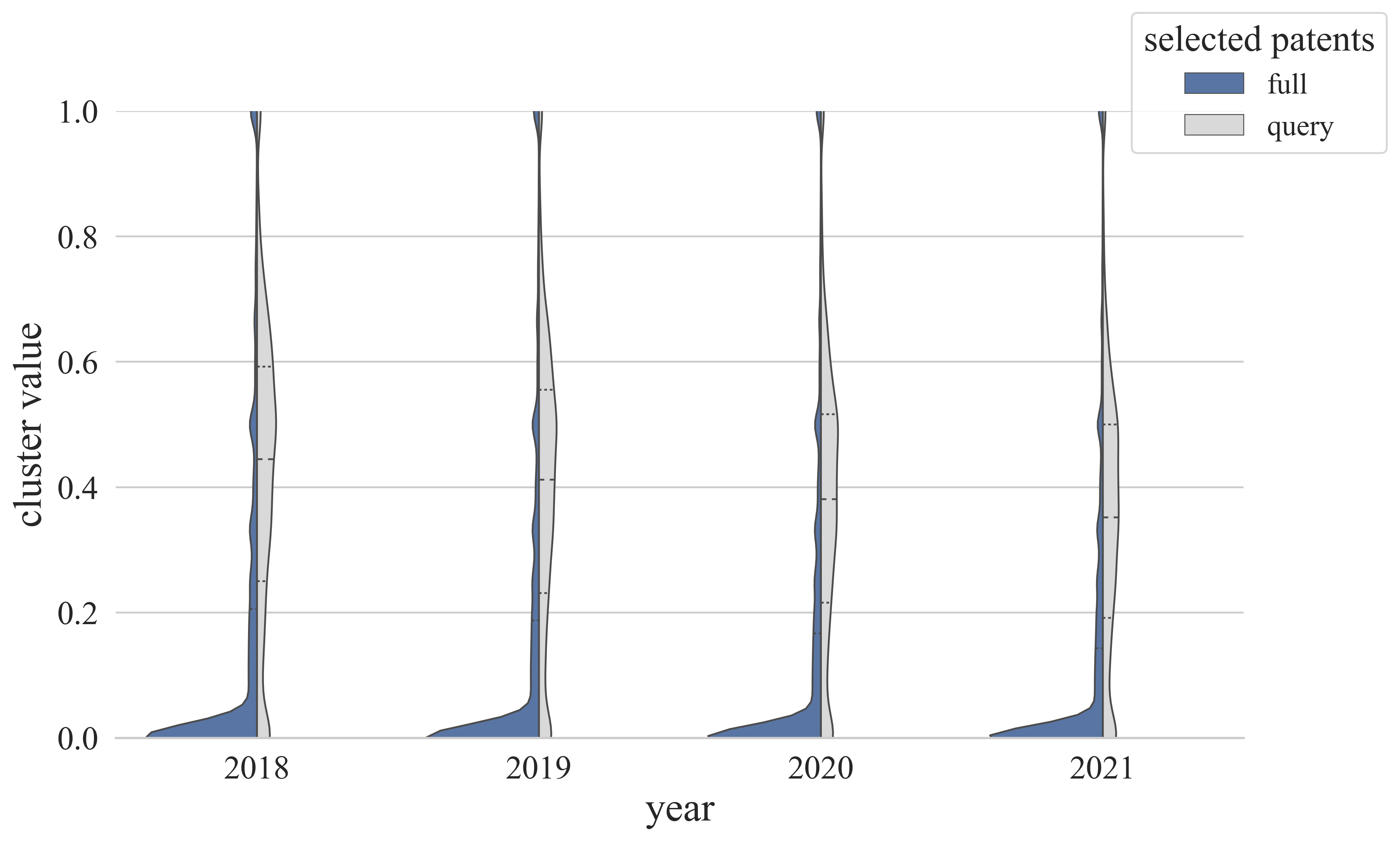}
    \caption{Violin plots for cluster value, from 2018 to 2021. The dashed lines represent the separations between the four quartiles. We see that clusters linked to cybersecurity are on average much more valuable than the average patent. This means the former patents receive more citations and attention than the latter.}
    \label{fig5}
\end{figure}

Figures \ref{fig3:a}, \ref{fig3:b}, \ref{fig4} and \ref{fig5} present the distribution of cluster size and cluster value for all technologies linked to our cybersecurity-related query. In Figure \ref{fig3:b}, the cumulative distribution function for cluster size shows that most cybersecurity technologies in the CPC topology have gathered fewer than 500 patents since their inception. Figure \ref{fig3:a} however shows that many cybersecurity-related technologies contain highly-valued patents. This becomes even more apparent when comparing these measures with statistics describing the entire patent database. This comparison, shown in Figures \ref{fig4} and \ref{fig5}, highlights the quality of cybersecurity-related patents. Technologies linked to cybersecurity show median cluster values between 0.3 and 0.5 for the last four years. Although the trend of cluster value points downward, the contrast to the values for the full data set is very evident; cybersecurity innovations are highly cited compared to other patents.

Next, we present the technology and company rankings of our recommender system. This recommender system is built atop the cybersecurity graph discussed in Chapter 4. Table \ref{Table7} in the Annex presents the 10 most highly-ranked CPC subgroup nodes in this graph based on their Technology Index. From these results, we notice a strong interest in biology-inspired computer systems (mainly neural networks), payment systems, WiFi \& control unit security, and cryptography.

We do not present here the ranking of the Assignee Quality Index, because many companies therein have the maximum value of 1 and presenting a list of the top 10 companies in that table is futile; rather, that ranking is best used for looking up a company-specific value. In Table \ref{Table4}, we present the top 10 companies according to the Impact Index. Unsurprisingly, major software and hardware companies top the list, propelled mainly by their large patent count. Table \ref{Table5} scales the Impact Index based on each company's patent output size, and thus we can highlight smaller, lesser known companies that have a considerable impact on the cybersecurity market proportional to their size. We have only selected companies with at least 5 granted patents in the last year; this step filters out outlier companies that produced very few high-value patents and thus found themselves at the top of the list, more out of luck than the product of high-quality research that can be reproduced repeatedly over several years. Finally, Table \ref{Table6} shows the most influential companies in the cybersecurity market. We have again very dominant technology companies topping out the list, however hardware companies have the edge here. This could be explained by the business model of some of these companies, which integrate many software and hardware solutions into their customer products and hence must stay up-to-date in many cybersecurity-relevant topics.

\begin{table*}
    \footnotesize
    \centering
    \begin{tabular}{p{0.04\textwidth}p{0.52\textwidth}p{0.1\textwidth}p{0.15\textwidth}p{0.15\textwidth}}
    \toprule
    \textbf{\#} & \textbf{Company} & \textbf{Patent Count} & \textbf{Value} & \textbf{Impact} \\
    \midrule
    1  & International Business Machines Corporation    & 7314  & 0.479355  & 3,488,180 \\
    2  & Microsoft Technology Licensing, LLC            & 2644  & 0.815053  & 3,157,710 \\
    3  & Amazon Technologies, Inc.                      & 2133  & 0.487107  & 1,632,140 \\
    4  & Cisco Technology, Inc                          & 1040  & 0.779808  & 1,532,470 \\
    5  & Advanced New Technologies Co, Ltd              & 518   & 0.486486  & 1,453,670 \\
    6  & Intel Corporation                              & 2788  & 0.457317  & 1,283,470 \\
    7  & EMC IP Holding Company LLC                     & 1235  & 0.733603  & 1,265,040\\
    8  & Apple Inc.                                     & 2568  & 0.54595   & 1,132,630 \\
    9  & AS America, Inc.                               & 496   & 0.59879   & 941,869 \\
    10 & Google LLC                                     & 1621  & 0.676126  & 921,388 \\
    \bottomrule
    \end{tabular}
    \vspace{0.5cm}
    \caption{Companies ranked according to the Impact Index. The list is made up of major software and hardware companies, boosted by a large number of impactful patents.}
    \label{Table4}
\end{table*}

\begin{table*}
    \footnotesize
    \centering
    \begin{tabular}{p{0.04\textwidth}p{0.50\textwidth}p{0.1\textwidth}p{0.10\textwidth}p{0.12\textwidth}}
    \toprule
    \textbf{\#} & \textbf{Company} & \textbf{Value} & \textbf{Patent Count} & \textbf{Norm. Impact}\\
    \midrule
        1   & CyberArk Software Ltd.                & 0.7567    & 37 & 833.551 \\ 
        2   & Intertrust Technologies Corporation   & 0.8571    & 14 & 771.041 \\
        3   & SHAPE SECURITY, INC.                  & 0.9090    & 11 & 769.185 \\ 
        4   & F5 NETWORKS, INC.                     & 0.9230    & 26 & 765.03  \\ 
        5   & Sophos Limited                        & 0.8695    & 46 & 744.879 \\ 
        6   & McAfee, LLC                           & 0.8924    & 93 & 730.186 \\ 
        7   & SONICWALL INC.                        & 0.8571    & 14 & 726.287 \\ 
        8   & MX TECHNOLOGIES, INC.                 & 0.875     & 16 & 717.432 \\ 
        9   & FireEye, Inc.                         & 0.9787    & 47 & 683.293 \\ 
        10  & Netskope, Inc.                        & 0.75      & 20 & 632.204 \\ 
    \bottomrule
    \end{tabular}
    \vspace{0.5cm}
    \caption{Companies ranked according to the Normalised Impact Index. Companies with fewer than 5 patents are removed for more useful results. This ranking highlights small but impactful companies.}
    \label{Table5}
\end{table*}

\begin{table*}
    \footnotesize
    \centering
    \begin{tabular}{p{0.04\textwidth}p{0.5\textwidth}p{0.1\textwidth}p{0.15\textwidth}p{0.1\textwidth}}
    \toprule
    \textbf{\#} & \textbf{Company} & \textbf{Patent Count} & \textbf{Value} & \textbf{Influence}\\
    \midrule
    1  & International Business Machines Corporation        & 7314  & 0.4793 & 0.3573 \\
        2  & Samsung Electronics Co., Ltd.                  & 5415  & 0.3566 & 0.2811 \\
    3  & Qualcomm Incorporated                              & 2129  & 0.5758 & 0.1948 \\
    4  & Huawei Technologies Co., Ltd.                      & 2765  & 0.0239 & 0.1787 \\
    5  & LG Electronics, Inc.                               & 2094  & 0.1905 & 0.1763 \\
    6  & Apple Inc.                                         & 2568  & 0.5459 & 0.1668 \\
    7  & Microsoft Technology Licensing, LLC.               & 2644  & 0.8150 & 0.1608 \\
    8  & Intel Corporation                                  & 2788  & 0.4573 & 0.1599 \\
    9  & Amazon Technologies, Inc.                          & 2133  & 0.4871 & 0.106 \\
    10 & Facebook, Inc.                                     & 1317  & 0.2498 & 0.1025 \\
    \bottomrule
    \end{tabular}
    \vspace{0.5cm}
    \caption{Companies ranked according to the Influence Index. Many of these assignees are hardware companies with ties to a large set of technologies.}
    \label{Table6}
\end{table*}

\begin{figure}
    \centering
    \figuretitle{Assignee patent count vs Normalised impact}
    \includegraphics[width=0.8\textwidth]{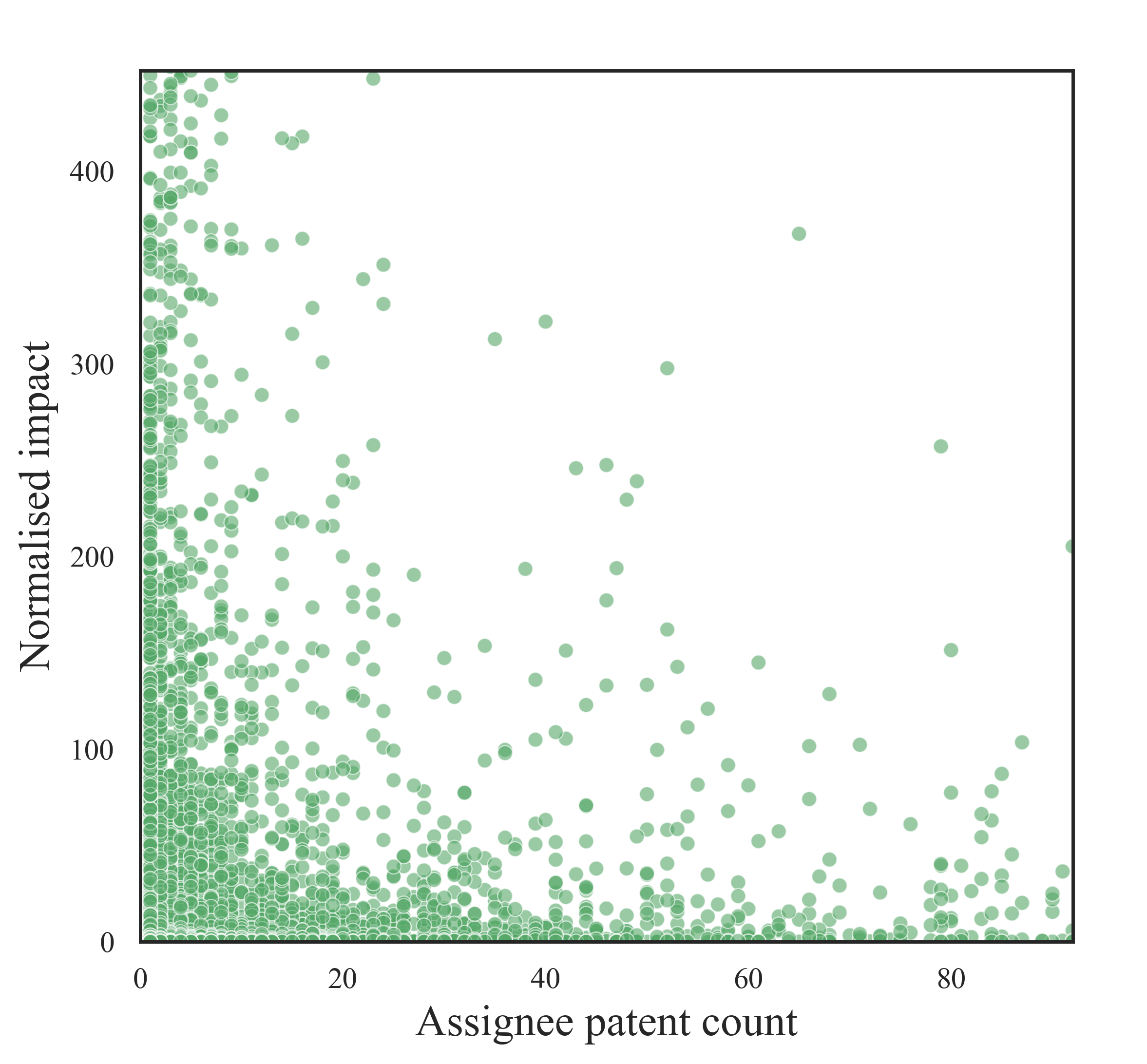}
    \caption{Plot of patent count against the normalised impact index, for all topical clusters. It seems that both variables are correlated through a negative exponential, which supports the economic theory of decreasing marginal returns in large companies. Further analysis could confirm this link.}
    \label{fig6}
\end{figure}

\begin{figure}[!h]
    \centering
    \figuretitle{Assignee influence vs Assignee impact}
    \includegraphics[width=0.8\textwidth]{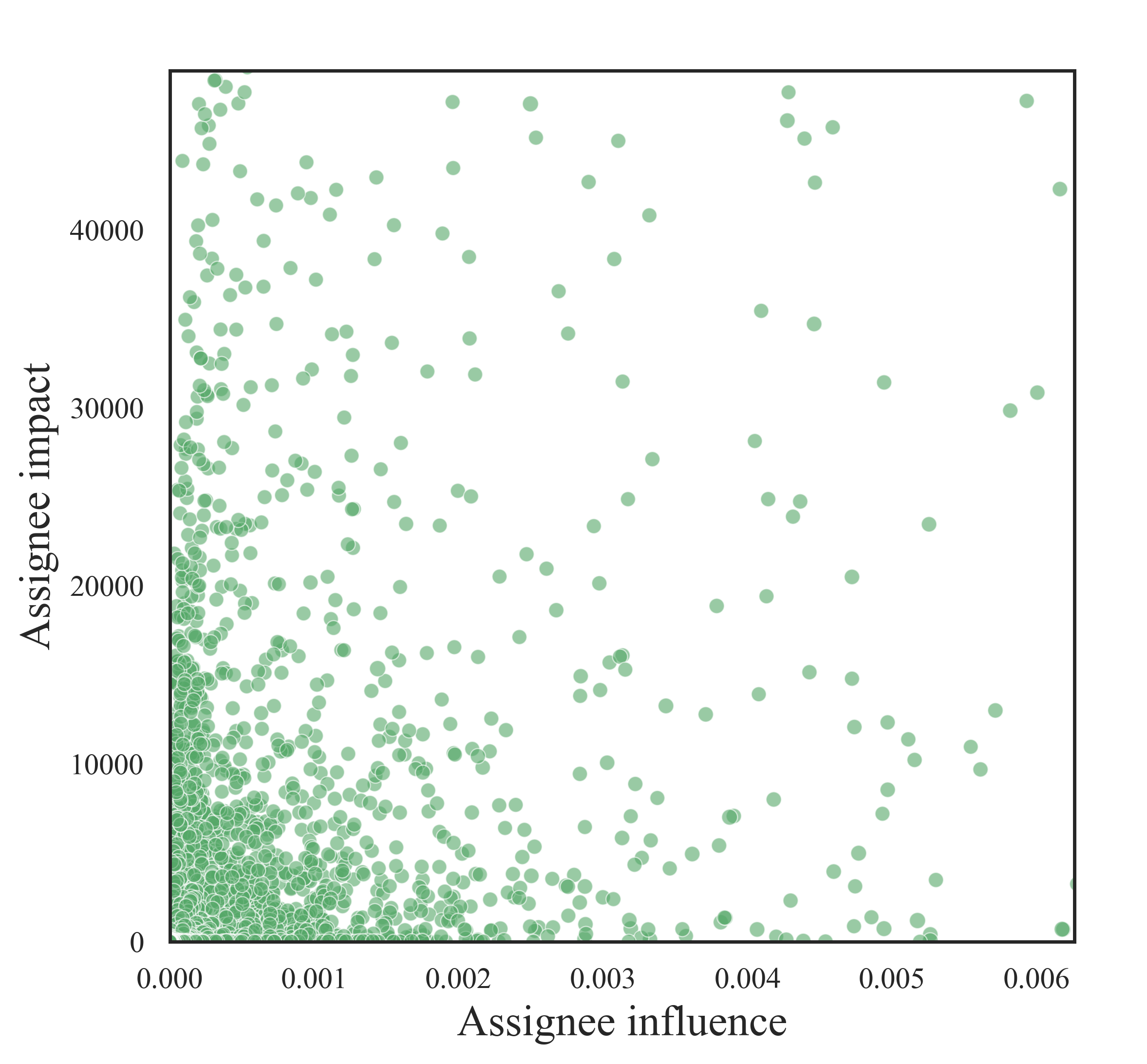}
    \caption{Plot of cluster influence against cluster impact, for all topical clusters. There is no apparent relationship between the two.}
    \label{fig7}
\end{figure}

Moreover, by cross-plotting several of these statistics, we can glean additional insights for research and development (R\&D) management. Figure \ref{fig6} plots for each assignee (i) the patent count against (ii) the normalised impact of the company on the cybersecurity market. Based on preliminary analysis, it seems that the two indicators are negatively correlated. This supports the economic theory of decreasing marginal return in large companies; the larger the R\&D output of technology companies, the less valuable each cybersecurity innovation seems to be. Figure \ref{fig7} in the Annex plots assignee influence against assignee impact, with no statistically significant trend. It is therefore not possible to state that horizontal integration or investments into many corners of the cybersecurity market improve a company's impact within the field. 

\section{Discussion}

 Each indicator in our case study serves different cybersecurity and cyberdefence end users. The Technology Index allows companies and governments to allocate their resources to rising and valuable technologies in the cyberspace. It also hinders blindspots in an institution's cyber threat management, by highlighting all technologies and areas of business where cybersecurity plays an important role. The Technology Index can also help steer industrial policy and encourage early adoption of industry standards for emerging technologies.

The Impact Index can be used to identify leading companies in cybersecurity. These companies set technology standards and employ the largest share of the field's experts. Economists therefore have an incentive to follow the progress of these major market players. Thanks to the Impact Index, cybersecurity engineers and consultants also know where to purchase the most established and well-funded cybersecurity products. The Normalised Impact index on the other hand can be useful for early-stage investments into promising cybersecurity firms with the most near-term potential. Finally, the Influence Index allows clients to partner up with the best-connected companies in the cybersecurity market, either to explore the market at large, jointly develop additional products or to hire consultancy services from.

The following description could be a potential use-case of these indicators. By cross-correlating the Impact Index and the Technology Index, a national government can spot strong and weak areas of the local cybersecurity market. Using the results of the Normalised Impact Index, different ministries can enact targeted investments into the most promising native startups. This analysis is also bolstered by an API built into the recommender system, which indexes the most important technologies for a selected company or the most prolific companies for a selected technology (see Figure \ref{fig2}).
\section{Conclusion, Limitations, Future scope}
In order to investigate the cybersecurity technology and market landscape, we have built a generalisable and lightweight patent-analysis system. First, with the help of a best-in-class machine learning framework, we split patents into high- and low-value classes using patent metadata as training inputs. From this binary classification and with the help of patent count time-series, we derived the growth in interest for all technologies of the CPC classification scheme. Furthermore, we measured various company scores based on their patent portfolio. Running this system on the cybersecurity industry, our results show highly satisfactory rankings of emerging technologies and companies, with outputs corresponding to expert expectations for such a system. 
We also recognise the need for further research to look beyond patents as a source of information. Indeed, some technology fields experience more pressure to submit patent applications, and this is especially true in long-standing technology companies such as IBM. These differences naturally skew our system output towards companies and technology fields that make patent submissions a priority. Given the current advances in Natural Language Processing and computational Bayesian methods, we also argue that recognising topic-related patents could be improved with state-of-the-art topic extraction and modelling methods. Overall, we believe our system to be a first stepping stone towards a highly intelligent market and technology forecasting tool.

\clearpage 

\printbibliography

@article{Small2014,
  title =	 "Identifying emerging topics in science and
                  technology",
  journal =	 "Research Policy",
  volume =	 "43",
  number =	 "8",
  pages =	 "1450 - 1467",
  year =	 "2014",
  issn =	 "0048-7333",
  doi =		 "https://doi.org/10.1016/j.respol.2014.02.005",
  url =
                  "http://www.sciencedirect.com/science/article/pii/S0048733314000298",
  author =	 "Henry Small and Kevin W. Boyack and Richard Klavans"
}

@article{Martin1995,
  author =	 { Ben R. Martin },
  title =	 {Foresight in science and technology},
  journal =	 {Technology Analysis \& Strategic Management},
  volume =	 {7},
  number =	 {2},
  pages =	 {139-168},
  year =	 {1995},
  publisher =	 {Routledge},
  doi =		 {10.1080/09537329508524202},
  URL =		 { https://doi.org/10.1080/09537329508524202 },
  eprint =	 { https://doi.org/10.1080/09537329508524202 }
}

@TECHREPORT{Corrocher2003,
  title =	 {The emergence of new technologies in the ICT field:
                  main actors, geographical distribution and knowledge
                  sources},
  author =	 {Corrocher, Nicoletta and Malerba, Franco and
                  Montobbio, Fabio},
  year =	 {2003},
  institution =	 {Department of Economics, University of Insubria},
  type =	 {Economics and Quantitative Methods},
  url =
                  {https://EconPapers.repec.org/RePEc:ins:quaeco:qf0317}
}

@article{Hung2006,
  title =	 "Stimulating new industries from emerging
                  technologies: challenges for the public sector",
  journal =	 "Technovation",
  volume =	 "26",
  number =	 "1",
  pages =	 "104 - 110",
  year =	 "2006",
  issn =	 "0166-4972",
  doi =		 "https://doi.org/10.1016/j.technovation.2004.07.018",
  url =
                  "http://www.sciencedirect.com/science/article/pii/S0166497204001324",
  author =	 "Shih-Chang Hung and Yee-Yeen Chu"
}

@article{Boon2008,
  title =	 "Exploring emerging technologies using metaphors: A
                  study of orphan drugs and pharmacogenomics",
  journal =	 "Social Science \& Medicine",
  volume =	 "66",
  number =	 "9",
  pages =	 "1915 - 1927",
  year =	 "2008",
  issn =	 "0277-9536",
  doi =		 "https://doi.org/10.1016/j.socscimed.2008.01.012",
  url =
                  "http://www.sciencedirect.com/science/article/pii/S0277953608000385",
  author =	 "Wouter Boon and Ellen Moors"
}

@article{Halaweh2013,
  author =	 {Mohanad Halaweh},
  title =	 {Emerging Technology: What is it?},
  journal =	 {Journal of Technology Management \& Innovation},
  volume =	 {8},
  number =	 {3},
  year =	 {2013},
  issn =	 {0718-2724},
  pages =	 {108--115},
  doi =		 {10.4067/S0718-27242013000400010},
  url =
                  {http://www.jotmi.org/index.php/GT/article/view/art400}
}

@article{Intepe2012, 
  author =	 {Gizem Intepe and Tufan Koc},
  title =	 {The Use of S Curves in Technology Forecasting and
                  its Application On 3D TV Technology},
  journal =	 {International Journal of Industrial and
                  Manufacturing Engineering},
  volume =	 {6},
  number =	 {11},
  year =	 {2012}
}

@article{Nieto1998,
  author =	 "Mariano Nieto and Francisco Lop\'{e}z and Fernando
                  Cruz",
  title =	 "Performance analysis of technology using the S curve
                  model: the case of digital signal processing (DSP)
                  technologies",
  journal =	 "Technovation",
  volume =	 "18",
  number =	 "6",
  pages =	 "439 - 457",
  year =	 "1998",
  issn =	 "0166-4972",
  doi =
                  "https://dRotolo2015oi.org/10.1016/S0166-4972(98)00021-2",
  url =
                  "http://www.sciencedirect.com/science/article/pii/S0166497298000212"
}

@inproceedings{Zhou2018,
  title =	 {Forecasting emerging technologies with deep learning
                  and data augmentation: convergence emerging
                  technologies vs non-convergence emerging
                  technologies},
  author =	 {Yuan Zhou and Fang Dong and Zhaofu Li and JunFei Du
                  and Yufei Liu and Li Zhang},
  year =	 {2018},
  booktitle =	 {Proceedings of 6th International Conference on
                  Future-Oriented Technology Analysis (FTA)},
  URL =
                  {https://ec.europa.eu/jrc/sites/jrcsh/files/fta2018-paper-b3-zhou.pdf},
}

@article{Kyebambe2017,
  title =	 "Forecasting emerging technologies: A supervised
                  learning approach through patent analysis",
  journal =	 "Technological Forecasting and Social Change",
  volume =	 "125",
  pages =	 "236 - 244",
  year =	 "2017",
  issn =	 "0040-1625",
  doi =		 "https://doi.org/10.1016/j.techfore.2017.08.002",
  url =
                  "http://www.sciencedirect.com/science/article/pii/S0040162516307065",
  author =	 "Moses Ntanda Kyebambe and Ge Cheng and Yunqing Huang
                  and Chunhui He and Zhenyu Zhang"
}

@article{Daim2006,
  title =	 "Forecasting emerging technologies: Use of
                  bibliometrics and patent analysis",
  journal =	 "Technological Forecasting and Social Change",
  volume =	 "73",
  number =	 "8",
  pages =	 "981 - 1012",
  year =	 "2006",
  note =	 "Tech Mining: Exploiting Science and Technology
                  Information Resources",
  issn =	 "0040-1625",
  doi =		 "https://doi.org/10.1016/j.techfore.2006.04.004",
  url =
                  "http://www.sciencedirect.com/science/article/pii/S0040162506001168",
  author =	 "Tugrul U. Daim and Guillermo Rueda and Hilary Martin
                  and Pisek Gerdsri"
}

@inproceedings{Ranaei2014,
  Author =	 {S. Ranaei and M. Karvonen and A. Suominen and
                  T. K{\"a}ssi},
  Booktitle =	 {Proceedings of PICMET '14 Conference: Portland
                  International Center for Management of Engineering
                  and Technology; Infrastructure and Service
                  Integration},
  Journal =	 {Proceedings of PICMET '14 Conference: Portland
                  International Center for Management of Engineering
                  and Technology; Infrastructure and Service
                  Integration},
  Pages =	 {2924--2937},
  Title =	 {Forecasting emerging technologies of low emission
                  vehicle},
  Ty =		 {CONF},
  Year =	 {2014},
}

@inproceedings{Kucharavy2009,
  author =	 {D. Kucharavy and E. Schenk and R. De Guio},
  title =	 {Long-Run Forecasting of Emerging Technologies with
                  Logistic Models and Growth of Knowledge},
  booktitle =	 {Proceedings of the 19th CIRP Design Conference -
                  Competitive Design, Cranfield University, 30-31
                  March, 2009},
  year =	 {2009},
  pages =	 {277}
}

@article{Bengisu2006,
  title =	 "Forecasting emerging technologies with the aid of
                  science and technology databases",
  journal =	 "Technological Forecasting and Social Change",
  volume =	 "73",
  number =	 "7",
  pages =	 "835 - 844",
  year =	 "2006",
  issn =	 "0040-1625",
  doi =		 "https://doi.org/10.1016/j.techfore.2005.09.001",
  url =
                  "http://www.sciencedirect.com/science/article/pii/S0040162505001393",
  author =	 "Murat Bengisu and Ramzi Nekhili"
}

@article{andersen1999,
  title={The hunt for S-shaped growth paths in technological innovation: a patent study},
  author={Andersen, Birgitte},
  journal={Journal of evolutionary economics},
  volume={9},
  number={4},
  pages={487--526},
  year={1999},
  publisher={Springer}
}

@article{haupt2007patent,
  title={Patent indicators for the technology life cycle development},
  author={Haupt, Reinhard and Kloyer, Martin and Lange, Marcus},
  journal={Research Policy},
  volume={36},
  number={3},
  pages={387--398},
  year={2007}, 
  publisher={Elsevier}
}

@article{meyer2001patent,
  title={Patent citation analysis in a novel field of technology: An exploration of nano-science and nano-technology},
  author={Meyer, Martin},
  journal={Scientometrics},
  volume={51},
  number={1},
  pages={163--183},
  year={2001},
  publisher={Akad{\'e}miai Kiad{\'o}, co-published with Springer Science+ Business Media BV, Formerly Kluwer Academic Publishers BV}
}

@article{porter2002,
  title={Measuring national ‘emerging technology’capabilities},
  author={Porter, Alan L and Roessner, J David and Jin, Xiao-Yin and Newman, Nils C},
  journal={Science and Public Policy},
  volume={29},
  number={3},
  pages={189--200},
  year={2002},
  publisher={Beech Tree Publishing}
}

@article{Lee2018,
  title =	 "Early identification of emerging technologies: A machine learning approach using multiple patent indicators",
  author =	 "Lee, Changyong and Kwon, Ohnjin and Kim, Myeongjung an Kwon, Daeil",
  year =	 "2018",
  doi =		 "10.1016/j.techfore.2017.10.002",
  language =	 "English",
  volume =	 "127",
  pages =	 "291--303",
  journal =	 "Technological Forecasting and Social Change",
  issn =	 "0040-1625",
  publisher =	 "Elsevier Inc.",
  number =	 "3",
}

@article{mezzetti2021,
  title =	 "TechRank: A Network-Centrality Approach for Informed Cybersecurity-Investment",
  author =	 "Mezzetti, Anita and Percia David, Dimitri and Maillart, Thomas and Tsesmelis, Michael and Mermoud, Alain",
  year =	 "2021",
  doi =		 "arXiv:2112.05548",
  language =	 "English",
  pages =	 "1--7",
  publisher =	 "Arxiv",
}

@article{makridakis_1983,
  title =	 "Averages of Forecasts: Some Empirical Results",
  author =	 "Makridakis, Spyros and Winkler, Robert L.",
  year =	 "1983",
  doi =		 "10.1287/mnsc.29.9.987",
  language =	 "English",
  volume =	 "29",
  pages =	 "987--996",
  journal =	 "Management Science",
  number =	 "9",
}

@article{wang_2019,
  title =	 "An approach to identify emergent topics of technological convergence: A case study for 3D printing",
  author =	 "Wang, Zhinan and Porter, Alan and Wang, Xuefeng and Carley, Stephen",
  year =	 "2019",
  doi =		 "10.1016/j.techfore.2018.12.015",
  language =	 "English",
  volume =	 "146",
  pages =	 "723--732",
  journal =	 "Technological Forecasting and Social Change",
}

@article{clemen_1989,
  title =	 "Combining forecasts: A review and annotated bibliography",
  author =	 "Clemen, Robert T.",
  year =	 "1989",
  doi =		 "10.1016/0169-2070(89)90012-5",
  language =	 "English",
  volume =	 "5",
  pages =	 "559--583",
  journal =	 "International Journal of Forecasting",
  publisher =	 "Elsevier",
  number =	 "4",
}

@article{Ruhnau_2000,
  title =	 "Eigenvector-centrality - a node-centrality?",
  author =	 "Ruhnau, Britta",
  year =	 "2000",
  doi =		 "10.1016/S0378-8733(00)00031-9",
  language =	 "English",
  volume =	 "22",
  pages =	 "357--365",
  journal =	 "Social Networks",
  publisher =	 "Elsevier",
  number =	 "4",
}

@misc{patentsview, title={PatentsView}, url={http://www.patentsview.org/}, journal={PatentsView}, publisher={Patentsview}}

\section{Annex}
\label{section:annex}

\begin{table*}[h!]
    \footnotesize
    \centering
    \begin{tabular}{p{0.09\textwidth}p{0.16\textwidth}p{0.1\textwidth}p{0.14\textwidth}p{0.51\textwidth}}
    \toprule
    \textbf{Tech. Index} & \textbf{Subgroup} & \textbf{Patent count} & \textbf{Citation count} & \textbf{Description}\\
    \midrule
    
    8.02 & G06N3/08    & 6519  & 251.6      & Computer systems based on biological models-using neural network models-Learning methods\\ 
    8.02 & G06F7/50    & 163   & 13.7       & Methods or arrangements for processing data by operating upon the order or content of the data handled -Methods or arrangements for performing computations using exclusively denominational number representation, e.g. using binary, ternary, decimal representation-using non-contact-making devices, e.g. tube, solid state device; using unspecified devices-Adding; Subtracting\\
    7.92 & G06Q20/3558 & 132   & 81.4       & Payment architectures, schemes or protocols -characterised by the use of specific devices ; or networks-using cards, e.g. integrated circuit [IC] cards or magnetic cards-Personalisation of cards for use-Preliminary personalisation for transfer to user\\
    7.90 & G06N3/0454  & 5458  & 141.3      & Computer systems based on biological models-using neural network models-Architectures, e.g. interconnection topology-using a combination of multiple neural nets\\ 
    7.55 & G06N20/00   & 13947 & 747.4      & Machine learning \\
    7.48 & H04W80/08   & 279   & 14.6       & Wireless network protocols or protocol adaptations to wireless operation-Upper layer protocols\\
    7.15 & G11C8/20    & 243   & 53.7       & Arrangements for selecting an address in a digital store -Address safety or protection circuits, i.e. arrangements for preventing unauthorized or accidental access \\ 
    7.06 & G06F9/3818  & 66    & 61.4       & Arrangements for program control, e.g. control units -using stored programs, i.e. using an internal store of processing equipment to receive or retain programs-Arrangements for executing machine instructions, e.g. instruction decode -Concurrent instruction execution, e.g. pipeline, look ahead-Decoding for concurrent execution \\ 
    7.05 & H04L2209/38 & 2858  & 2728.0     & Additional information or applications relating to cryptographic mechanisms or cryptographic arrangements for secret or secure communication H04L9/00-Chaining, e.g. hash chain or certificate chain \\ 
    6.91 & G06N3/0472 & 977   & 51.0        & Computer systems based on biological models-using neural network models-Architectures, e.g. interconnection topology-using probabilistic elements, e.g. p-rams, stochastic processors\\ 
    \bottomrule
    \end{tabular}
    \vspace{0.5cm}
    \caption{Technologies ranked according to Technology Index. Important emerging technologies in cybersecurity are linked to neural networks, payment systems, WiFi, control units and cryptography.}
    \label{Table7}
\end{table*}

\end{document}